\documentclass[aps,prd,twocolumn,showpacs,amsmath,amssymb]{revtex4-1}
\usepackage{amsmath}
\usepackage{graphicx}
\usepackage{subfigure}
\usepackage{epstopdf}
\usepackage{color}
\usepackage{multirow}
\usepackage{setspace}
\usepackage{overpic}
\usepackage{amssymb}
\usepackage{lineno}
\usepackage{bm}
\usepackage{rotating}
\usepackage{makecell}
\usepackage[utf8]{inputenc}
\usepackage{hyperref}
\hypersetup{
	colorlinks=true,
	linkcolor=blue,
	filecolor=blue,
	urlcolor=blue,
	citecolor=blue,
}
\let\oldequation\equation
\let\oldendequation\endequation

\renewenvironment{equation}
  {\linenomathNonumbers\oldequation}
  {\oldendequation\endlinenomath}

\begin{document}

\title{\bf \boldmath
Precision Measurement of the Branching Fraction of $D^{+}\to \mu^{+}\nu_{\mu}$}

\author{
M.~Ablikim$^{1}$, M.~N.~Achasov$^{4,c}$, P.~Adlarson$^{76}$, O.~Afedulidis$^{3}$, X.~C.~Ai$^{81}$, R.~Aliberti$^{35}$, A.~Amoroso$^{75A,75C}$, Q.~An$^{72,58,a}$, Y.~Bai$^{57}$, O.~Bakina$^{36}$
% ...
\\
\vspace{0.2cm}
(BESIII Collaboration)\\
\vspace{0.2cm} {\it
$^{1}$ Institute of High Energy Physics, Beijing 100049, People's Republic of China\\
% ...
$^{3}$ Bochum Ruhr-University, D-44780 Bochum, Germany\\
$^{4}$ Budker Institute of Nuclear Physics SB RAS (BINP), Novosibirsk 630090, Russia\\
% ...
$^{35}$ Johannes Gutenberg University of Mainz, Johann-Joachim-Becher-Weg 45, D-55099 Mainz, Germany\\
$^{36}$ Joint Institute for Nuclear Research, 141980 Dubna, Moscow region, Russia\\
% ...
$^{57}$ Southeast University, Nanjing 211100, People's Republic of China\\
$^{58}$ State Key Laboratory of Particle Detection and Electronics, Beijing 100049, Hefei 230026, People's Republic of China\\
% ...
$^{72}$ University of Science and Technology of China, Hefei 230026, People's Republic of China\\
% ...
$^{75}$ University of Turin and INFN, (A)University of Turin, I-10125, Turin, Italy; (B)University of Eastern Piedmont, I-15121, Alessandria, Italy; (C)INFN, I-10125, Turin, Italy\\
$^{76}$ Uppsala University, Box 516, SE-75120 Uppsala, Sweden\\
% ...
$^{81}$ Zhengzhou University, Zhengzhou 450001, People's Republic of China\\
\vspace{0.2cm}
$^{a}$ Deceased\\
% ...
$^{c}$ Also at the Novosibirsk State University, Novosibirsk, 630090, Russia\\
% ...
}
}

\begin{abstract}
Using $20.3~\mathrm{fb}^{-1}$ of $e^+e^-$ collision data collected at a center-of-mass energy of $E_{\rm c.m.}=3.773$ GeV with the BESIII detector operating at the BEPCII collider, we determine the branching fraction of the leptonic decay $D^+\to\mu^+\nu_\mu$
to be $(4.034\pm0.080_{\rm stat}\pm0.040_{\rm syst})\times10^{-4}$.
Interpreting our measurement with knowledge of the Fermi coupling constant $G_F$, the masses of the $D^+$ and $\mu^+$ as well as the lifetime of the $D^+$, we determine $f_{D^+}|V_{cd}|=(48.02\pm0.48_{\rm stat}\pm0.24_{\rm syst}\pm0.12_{\rm input}\pm0.15_{\rm EM})~\mathrm{MeV}$ after taking into account necessary radiative corrections. This result is  a factor of 2.3 more precise than the previous best measurement.
Using the value of the magnitude of the $c\to d$ Cabibbo-Kobayashi-Maskawa matrix element $|V_{cd}|$ given by the global standard model fit, we obtain the $D^+$ decay constant $f_{D^+}=(213.5\pm2.1_{\rm stat}\pm1.1_{\rm syst}\pm0.8_{\rm input}\pm0.7_{\rm EM})$\,MeV. Alternatively, using the value of $f_{D^+}$ from a precise lattice quantum chromodynamics calculation, we extract $|V_{cd}|=0.2265\pm0.0023_{\rm stat}\pm0.0011_{\rm syst}\pm0.0009_{\rm input}\pm0.0007_{\rm EM}$.
\end{abstract}

\maketitle

\oddsidemargin  -0.2cm
\evensidemargin -0.2cm

The leptonic decays of charmed mesons offer an important test-bed to access the quark mixing-matrix elements and test lepton flavor universality (LFU). In the standard model~(SM) of particle physics, the fully radiative inclusive decay rate of $D^+\to \ell^+\nu_\ell$ ($\ell=e$, $\mu$ or $\tau$) can be written as~\cite{decayrate}
\begin{equation}
\begin{split}
\Gamma_{D^+\to\ell^+\nu_\ell}&=\Gamma_{D^+\to\ell^+\nu_\ell}^{(0)}\left [1+\frac{\alpha}{\pi}C_p\right ]\\
&=\frac{G_F^2f^2_{D^+}m^3_{D^+}}{8\pi}|V_{cd}|^2\mu^2_{\ell}\left (1-\mu^2_{\ell} \right )^2\left [1+\frac{\alpha}{\pi}C_p\right ],
\label{eq1}
\end{split}
\end{equation}
where
$G_F$ is the Fermi coupling constant,
$f_{D^+}$ is the $D^+$ decay constant,
$|V_{cd}|$ is the magnitude of the $c\to d$
Cabibbo-Kobayashi-Maskawa~(CKM) matrix element,
$\mu_{\ell}$ is the ratio of the $\ell^+$ lepton mass to the $D^+$ meson mass ($m_{D^+}$),
and $[1+\frac{\alpha}{\pi}C_p]$ represents the radiative correction term~\cite{PDG2024}.
Previous measurements of $D^+\to \mu^+\nu_{\mu}$ have been performed by MARKIII~\cite{Adler:1987ty}, BES~\cite{BES:1998iue}, BESII~\cite{Tong:2005tr}, CLEO~\cite{CLEO:2004pwu,CLEO:2005jsh,CLEO:2008ffk}, and BESIII~\cite{BESIII:2013iro,BESIII:2019vhn,Ke:2023qzc} but with limited precision.
In contrast, the value of $f_{D^+}$ calculated by lattice quantum chromodynamics (LQCD)~\cite{Bazavov:2017lyh,FermilabLattice:2014tsy,Carrasco:2014poa,Boyle:2017jwu,Chen:2014hva,Dimopoulos:2013qfa,FlavourLatticeAveragingGroupFLAG:2021npn} has reached a precision of 0.3\%.
Precise measurements of $f_{D^+}$ and $|V_{cd}|$ are key to testing the LQCD calculations of $f_{D^+}$ and the CKM matrix unitarity at high precision.

By inputting $m_{\mu}$=105.6583755 MeV/$c^2$, $m_{\tau}$=1776.93 MeV/$c^2$, and $m_{D^+}$=1869.66 MeV/$c^2$~\cite{PDG2024} into $\Gamma_{D^+\to\ell^+\nu_\ell}^{(0)}$, the ratio of the branching fraction (BF) of $D^+\to\tau^+\nu_{\tau}$ to that of $D^+\to\mu^+\nu_{\mu}$ is expected to be $R_{\tau/\mu}=2.66\pm0.01$. Current experimental measurement~\cite{BESIII:2019vhn} is consistent with the SM prediction within the experimental uncertainty. However, there have been reports indication of potential LFU violation in semileptonic decays of $B$ mesons from the BaBar, LHCb, and Belle~\cite{PhysRevLett.109.101802,PhysRevD.88.072012,PhysRevLett.115.111803,PhysRevLett.113.151601,PhysRevLett.118.111801} experiments. 
Potential violation in (semi)leptonic $D$ decays may occur due to the interference between different amplitudes~\cite{PhysRevD.91.094009} or the interactions with scalar operators~\cite{Leng_2021}.  
Precision measurements of the BFs of $D^+\to\ell^+\nu_{\ell}$ offer an important test of LFU in the charm sector. 

This Letter reports a precise measurement of the BF of  $D^+\to\mu^+\nu_\mu$ obtained from the  analysis of 20.3~fb$^{-1}$ of $e^+e^-$
collision data collected in 2010, 2011, 2022, 2023 and 2024 with the BESIII detector at the center-of-mass energy of $E_{\rm c.m.} = 3.773$~GeV~\cite{Ablikim_2024}.
Charge-conjugate modes are always implied throughout this Letter unless stated specifically. The achieved precision is improved by a factor of 2.3 compared to
the previous best measurement~\cite{BESIII:2013iro}, which used 2.93 fb$^{-1}$ of data taken in 2010 and 2011.

The BESIII detector is a magnetic
spectrometer~\cite{BESIII:2009fln} located at the Beijing Electron
Positron Collider~\cite{Yu:IPAC2016-TUYA01}. The
cylindrical core of the BESIII detector consists of a helium-based
 multilayer drift chamber (MDC), a plastic scintillator time-of-flight
system (TOF), and a CsI (Tl) electromagnetic calorimeter (EMC),
which are all enclosed in a superconducting solenoidal magnet
providing a 1.0~T magnetic field. The solenoid is supported by an
octagonal flux-return yoke with resistive plate counter modules for muon identification interleaved with steel. More details of the design and performance of the BESIII detector can be found in
Ref.~\cite{BESIII:2009fln}. For 86\% of the data used in this Letter, the end-cap TOF was upgraded with multi-gap resistive plate chambers with a time resolution of
60\,ps~\cite{60ps1,60ps2}.  Simulated data samples produced with a {\sc
geant4}-based~\cite{geant4} Monte Carlo (MC) package, which
includes the geometric description of the BESIII detector and the
detector response, are used to determine detection efficiencies
and to estimate backgrounds. The beam-energy spread and initial-state radiation (ISR) in the $e^+e^-$
annihilations are simulated with the generator {\sc
kkmc}~\cite{kkmc2}. The inclusive MC sample includes the production of $D\bar{D}$
pairs (including quantum coherence for the neutral $D$ channels),
the non-$D\bar{D}$ decays of the $\psi(3770)$, the ISR
production of the $J/\psi$ and $\psi(3686)$ states, and the
continuum processes incorporated in {\sc kkmc}~\cite{kkmc,kkmc2}. All particle decays are modeled with {\sc evtgen}~\cite{evtgen,evtgen2} using BFs
either taken from the
Particle Data Group~\cite{PDG2024}, when available,
or otherwise estimated with {\sc lundcharm}~\cite{lundcharm,lundcharm2}.
Final-state radiation from charged final-state particles is incorporated using the {\sc
photos} package~\cite{photos}. The leptonic decay $D^{+}\to \mu^{+}\nu_{\mu}$ is simulated with {\sc PHOTOS  SLN} model~\cite{evtgen,evtgen2}.

At $E_{\rm c.m.}=3.773$\rm \,GeV, the $D^+D^-$ meson pairs are produced from $\psi(3770)$ decays
without accompanying hadrons. This favorable environment provides an ideal opportunity to study leptonic $D^+$ decays with the double-tag~(DT) method~\cite{MARK-III:1987jsm,ALBRECHT1990278}.
Initially, single-tag (ST) $D^-$ mesons are reconstructed via the eight hadronic decay modes
$K^{+}\pi^{-}\pi^{-}$, $K^0_{S}\pi^{-}$, $K^{+}\pi^{-}\pi^{-}\pi^{0}$, $K^0_{S}\pi^{-}\pi^{0}$,
$K^0_{S}\pi^{+}\pi^{-}\pi^{-}$, $K^{+}K^{-}\pi^{-}$, $\pi^{+}\pi^{-}\pi^{-}$, and $K^+\pi^{-}\pi^{-}\pi^{-}\pi^+$.
Then the $D^+\to\mu^+\nu_{\mu}$ candidates are selected by using the remaining tracks which have not been used in the selection of the tag side.
The event, in which the $D^+\to\mu^+\nu_{\mu}$ signal and the ST $D^-$ are simultaneously reconstructed, is called a DT event.
The BF of the $D^+\to\mu^+\nu_{\mu}$ decay is determined by
\begin{equation}
{\mathcal B}_{D^+\to\mu^+\nu_{\mu}} = \frac{N_{\rm DT}}{N^{\rm tot}_{\rm ST}\cdot \bar \epsilon_{\rm sig}},
\end{equation}
where $N_{\rm ST}^{\rm tot}$ is the total yield of ST $D^-$ mesons,
$N_{\rm DT}$ is the DT signal yield; and
$\bar \epsilon_{\rm sig}$ is the average signal efficiency weighted by the ST yields of the $i$-th tag mode in data,
   \begin{equation}
   {
\bar{{\mathcal \epsilon}}_{\rm sig} =\frac{\sum_i (N^i_{\rm ST}\cdot \epsilon^i_{\rm sig})}{N^{\rm tot}_{\rm ST}}=\frac{\sum_i (N^i_{\rm ST}\cdot \epsilon^i_{\rm DT}/\epsilon^i_{\rm ST})}{N^{\rm tot}_{\rm ST}},}
    \end{equation}
where $N^i_{\rm ST}$ is the number of ST $D^-$ mesons for the $i$-th tag mode in data, $\epsilon_{\rm sig}^{i}$ is the signal efficiency of the $i$-th tag mode,
$\epsilon^i_{\rm ST}$ is the efficiency of reconstructing the ST mode $i$ (called the ST efficiency),
and $\epsilon^i_{ \rm DT}$ is the efficiency of finding the tag mode $i$ and the $D^+\to\mu^+\nu_{\mu}$ decay simultaneously (called the DT efficiency).

For charged tracks, not originating from $K^0_S$ decays,
the polar angles with respect to the MDC $z$ axis ($\theta$) are required to satisfy $|\!\cos\theta|<0.93$.
In addition, the distance of closest approach to the interaction point (IP) must be less than 1\,cm in the  transverse plane, $|V_{xy}|$ and less than 10\,cm along the $z$-axis, $|V_{z}|$. The particle identification (PID) for charged particles combines measurements of the energy deposition in the MDC and the time-of-flight in the TOF to form likelihoods $\mathcal{L}(h)~(h=K,\pi)$ for each hadron $h$ hypothesis. The charged particles are assigned a particle type based on the  hypothesis with the higher likelihood.

Each $K_{S}^0$ candidate is reconstructed from two oppositely charged tracks satisfying $|V_{z}|<$ 20~cm.
The two charged tracks are assigned to be $\pi^+\pi^-$ without requiring any further PID criteria.
 They are constrained to originate from a common vertex, which is required to be displaced from the IP by a flight distance of at least twice the vertex resolution. The $\chi^2$ of the vertex fits (primary vertex fit and second vertex fit), where the number of degrees of freedom is 3, is required to be less than 100. This requirement retains 99\% of signals.
The invariant mass of the $\pi^+\pi^-$ pair is required to be within $(0.487,0.511)$~GeV/$c^2$~\cite{PhysRevD.109.072003} corresponding to $\pm 4\sigma$ around the $K^0_S$ nominal mass. 

The $\pi^0$ candidates are reconstructed via the dominant decay $\pi^0\to\gamma\gamma$. Candidates with both photons detected in the  end-cap EMC are rejected because of poor resolution. The photon candidates are identified using isolated showers in the EMC. The EMC time deviation from the event start time is required to be within [0,\,700]\,ns. The selection retains 99\% of reconstructed signal photons and removes 75\% of background energy depositions in the EMC. The energy deposition in the EMC is required to be greater than 25~MeV in the barrel region ($|\!\cos\theta|<0.80$) and 50~MeV in the end-cap region ($0.86<|\!\cos\theta|<0.92$).
To exclude showers that originate from charged tracks,
the angle subtended by the EMC shower and the position of the closest charged track at the EMC must be greater than $10^{\circ}$.
The $\pi^0$ candidates are required to have the  invariant mass of the $\gamma\gamma$ lying within $(0.115,\,0.150)$\,GeV$/c^{2}$, corresponding to $\pm 3\sigma$ around the $\pi^0$ nominal mass.
A mass-constrained~(1C) fit, where the number of degrees of freedom is 1, to the nominal $\pi^{0}$ mass~\cite{PDG2024} is imposed on the photon pair, to improve the momentum resolution. The $\chi^2$ of the 1C kinematic fit is required to be less than 50.
The four-momentum of the $\pi^0$ candidate updated by this kinematic fit is retained for the subsequent analysis.

\begin{figure}[htbp]\centering
\includegraphics[width=1.00\linewidth]{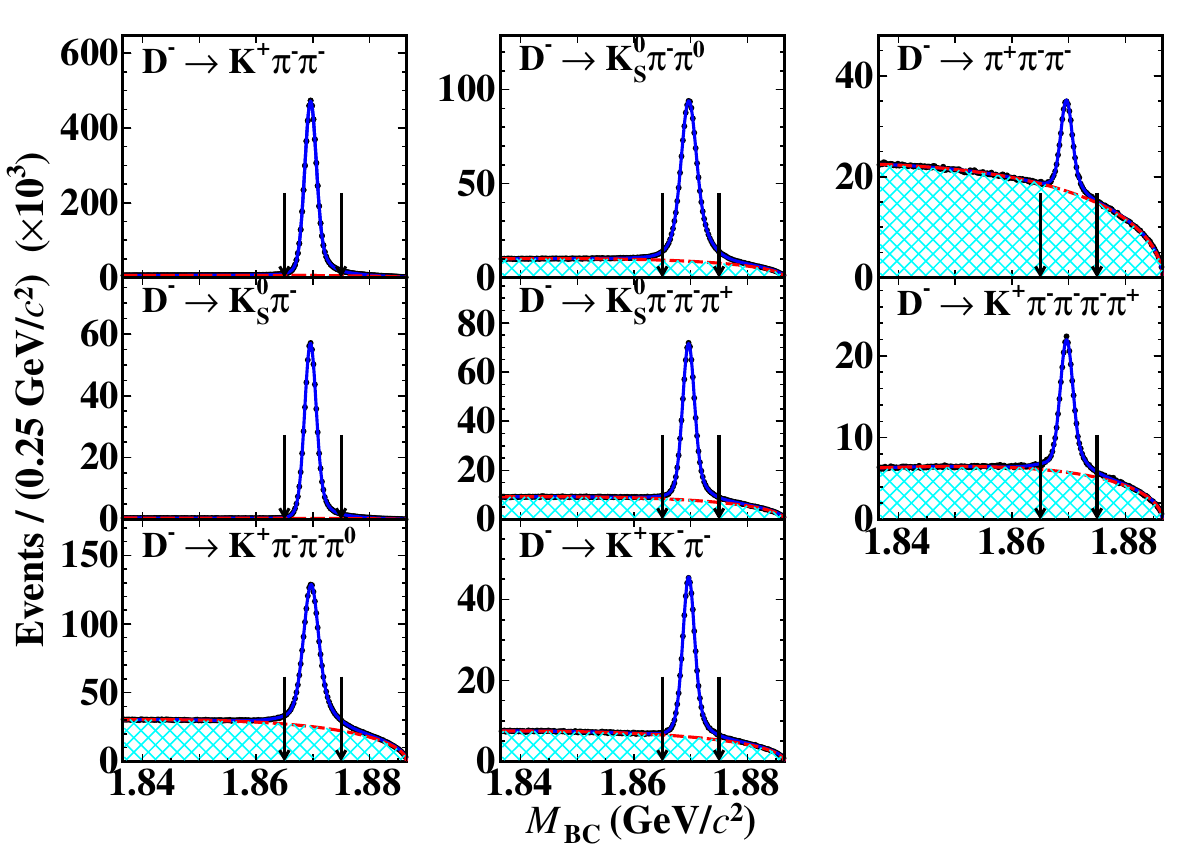}
\caption{Fits to the $M_{\rm BC}$ distributions of the
ST $D^-$ candidates. The dots with error bars are data. The blue solid curves are the fit results.
The red dashed curves are the fitted combinatorial backgrounds. The pairs of arrows denote the $M_{\rm BC}$ signal regions. The cyan hatched histograms are background events from the inclusive MC sample.}
\label{fig:datafit_Massbc}
\end{figure}

To separate the ST $D^-$ mesons from the combinatorial background, we define the energy difference $\Delta E\equiv E_{D^-}-E_{\mathrm{beam}}$ and the beam-constrained mass $M_{\rm BC}\equiv\sqrt{E_{\mathrm{beam}}^{2}-|\vec{p}_{D^-}|^{2}}$, where $E_{\mathrm{beam}}$ is the beam energy, and $E_{D^-}$ and $\vec{p}_{D^-}$ are the total energy and momentum of the ST $D^-$ meson in the $e^+e^-$ center-of-mass frame.
If there is more than one $D^-$ candidate in a given ST mode, the one with the smallest $|\Delta E|$ value is kept for the subsequent analysis.
The $\Delta E$ requirements and ST efficiencies are summarized in Table~\ref{ST:realdata}.

For each tag mode, the yield of ST $D^-$ mesons is extracted by fitting the corresponding $M_{\rm BC}$ distribution. In the fit, the signal shape is described as the sum of a simulated signal shape convolved with a double-normal distribution plus a single-normal distribution with free parameters. The double-normal and single-normal distributions account for different resolution and ISR effects between data and MC simulation, respectively.
The background shape is described by an ARGUS function~\cite{ALBRECHT1990278}, with the endpoint fixed at 1.8865~GeV/$c^{2}$ corresponding to $E_{\rm beam}$.
Figure~\ref{fig:datafit_Massbc} shows the results of the fits to the $M_{\rm BC}$ distributions of the accepted ST candidates for different tag modes in data. The candidates with $M_{\rm BC}$  lying  within $(1.865,1.875)$ GeV/$c^2$ are retained. We veto $D^-\to K_S^0\pi^-$ in $D^-\to\pi^+\pi^-\pi^-$ by requiring $|m_{\pi^+\pi^-}-0.4977|>0.03$ GeV/$c^2$. The contributions from the peaking backgrounds $D^-\to\pi^+\pi^-\pi^-, K_S^0e^-\bar{\nu}_e, K_S^0\mu^-\bar{\nu}_{\mu}$ in $D^-\to K_{S}^{0}\pi^{-}$, $D^-\to\pi^+\pi^+\pi^-\pi^-\pi^-$ in $D^{-} \to K_{S}^{0}\pi^{-}\pi^{-}\pi^{+}$, and $D^-\to K_S^0\pi^-$ in $D^-\to\pi^+\pi^-\pi^-$ are estimated by analyzing the inclusive MC sample and then are subtracted from the ST yields. These background fractions in the ST yields of  $D^-\to K_{S}^{0}\pi^{-}$, $D^{-} \to K_{S}^{0}\pi^{-}\pi^{-}\pi^{+}$, and $D^-\to\pi^+\pi^-\pi^-$ are 0.2\%, 0.1\%, and 2.8\%,  respectively. Summing all tag modes, we obtain a total yield of ST $D^-$ mesons of $(10789.0\pm3.9_{\rm stat})\times 10^3$.

\begin{table}
\renewcommand{\arraystretch}{1.2}
\centering
\caption {Requirements of $\Delta E$, ST $D^-$ yields in data, ST efficiencies ($\epsilon_{\rm ST}^{i}$), and DT efficiencies ($\epsilon^i_{\rm DT}$). The numbers in parentheses are the last two significant digits of the statistical uncertainties. The $\epsilon^i_{\rm ST}/\epsilon^i_{\rm DT}$ varies within 8\% for different tag modes, which are mainly caused by the significantly different signal
environments for some tag modes containing low momentum photon and pions in the signal and inclusive MC samples.}
\scalebox{0.87}{
\begin{tabular}{ccccc}
\hline
\hline
Tag mode                                                  & $\Delta E$~(MeV)     &  $N^{i}_{\rm ST}~(\times 10^3)$  &  $\epsilon^i_{\rm ST}~(\%)$ &  $\epsilon^i_{\rm DT}~(\%)$     \\\hline
$K^+\pi^-\pi^-$                                          &  $[-25,24]$       & $5449.1(25)$        &$50.1$          &$35.81(11)$\\
$K^{0}_{S}\pi^{-}$                                     &  $[-25,26]$       & $656.7 (08)$         &$50.6$          &$36.25(11)$\\
$K^{+}\pi^{-}\pi^{-}\pi^{0}$                         &  $[-57,46]$       & $1644.4 (18)$       &$23.3$          &$18.24(09)$\\
$K^{0}_{S}\pi^{-}\pi^{0}$                           &  $[-62,49]$        & $1384.8 (15)$      &$25.4$          &$19.33(09)$\\
$K^{0}_{S}\pi^{-}\pi^{-}\pi^{+}$                  &  $[-28,27]$        & $771.6 (11)$        &$28.9$           &$21.37(09)$\\
$K^{+}K^{-}\pi^{-}$                                    &  $[-24,23]$        & $472.6 (09)$        &$40.0$           &$28.88(10)$\\
$\pi^{+}\pi^{-}\pi^{-}$	                                &  $[-30,29]$	  & $204.7  (08)$      &$52.4$            &$37.77(11)$\\
$K^{+}\pi^{-}\pi^{-}\pi^{-}\pi^{+}$                &  $[-29,27]$	  & $215.1  (07)$      &$22.5$            &$16.69(08)$\\
\hline
\hline
          \end{tabular}
          }
\label{ST:realdata}
\end{table}

The $D^+\to\mu^+\nu_\mu$ candidates are selected in the  presence of the ST $D^-$ using the
remaining neutral clusters and charged tracks. The muon candidate must have an opposite charge
to the ST $D^-$ meson and
deposited energy within $(0.00,\,0.35)$\,GeV in the EMC. To separate muons from hadrons, requirements based on the muon hit depth ($d_{\mu^+}$) in the muon identifier modules are applied, taking into account the expected dependence on momentum ($p_{\mu^+}$) and flight direction $\cos\theta$.
These criteria are established from the distributions of $d_{\mu^+}$ versus $p_{\mu^+}$ using $e^+e^-\to(\gamma)\mu^+\mu^-$ candidates selected from data. The $|\!\cos\theta_{\mu^+}|$ and $p_{\mu^+}$ dependent requirements on $d_{\mu^+}$ follow those adopted in our previous measurements~\cite{bes33}.

To suppress backgrounds with extra photon(s),
the maximum energy of the unused showers in the DT
selection ($E^{\mathrm{extra}~\gamma}_{\rm max}$) is required to be less than 0.3\,GeV.
No additional charged track is allowed in the event. The yield of $D^+\to\mu^+\nu_\mu$  is determined by fitting the distribution of
the missing-mass squared of the undetected neutrino
\begin{linenomath}
\begin{align}
M_{\rm miss}^2&\equiv E_{\nu}^2-|\vec{p}_{\nu}|^2.
\end{align}
\end{linenomath}
Here $E_{\nu}\equiv E_{\rm c.m.}-E_{D^-}-E_{\mu}$ and $\vec p_{\nu}\equiv-\vec{p}_{D^-}-\vec{p}_{\mu}$, where $E_{\mu}$ and $\vec{p}_{\mu}$ denote
the energy and momentum of the muon, respectively.

The efficiencies of the DT reconstruction are determined with the signal MC samples, with $D^-$ decaying to tag modes and $D^+$ decaying to the signal mode.
Dividing these efficiencies by the ST efficiencies determined with the inclusive MC sample gives the corresponding
efficiencies of the $\mu^+\nu_\mu$ reconstruction.
The average efficiency over all tag modes
is determined to be $\bar{\epsilon}_{\rm sig}=(65.08\pm0.12)\%$.
This efficiency has been corrected by a factor of
\begin{equation}
f_{\mu\,\rm PID}^{\rm cor}=(88.7\pm0.1)\%, \nonumber
\end{equation}
to account for the differences of $\mu^+$ identification efficiencies between data and MC simulation, mainly due to the imperfect simulation of the $d_{\mu^+}$ variable~\cite{besiii2}. $f_{\mu\,\rm PID}^{\rm cor}$ is determined by using $e^+e^-\to\gamma\mu^+\mu^-$ events and reweighting by
the $\mu^+$ two-dimensional distribution in $|\!\cos\theta_{\mu^+}|$ and $p_{\mu^+}$ of $D^+\to\mu^+\nu_\mu$ decays.

The background includes two components. One consists of events with wrongly tagged $D^-$ decays (18.9\%), and the other contains correctly tagged $D^-$ decays but incorporating particle mis-identifications, which is mainly from the decays of $D^+\to \tau^+(\to \pi^+\bar\nu_\tau)\nu_\tau$ (4.8\%), $D^+\to\pi^+\pi^0$ (6.8\%), and $D^+\to\bar K^0\pi^+$ (30.2\%). These background fractions are counted over all backgrounds.
Analysis of the inclusive MC sample shows
that these two components make
comparable contributions and the main peaking backgrounds in the resulting $M_{\rm miss}^2$ distribution are $D^+\to \tau^+(\to \pi^+\bar\nu_\tau)\nu_\tau$ and $D^+\to\pi^+\pi^0$. Furthermore, the radiative decay $D^+\to\gamma\mu^+\nu_{\mu}$ from structure-dependent bremsstrahlung contribution, where the $D^+$ meson decays into a real photon and an off-shell vector meson (the meson subsequently decays weakly to $\mu^+$ and $\nu_{\mu}$), can also contribute a peaking structure in the resulting $M_{\rm miss}^2$ distribution. The yield of $D^+\to \gamma \mu^+\nu_\mu$ is estimated to be $6.6\pm4.6$ with the BF of $D^+\to\gamma e^+\nu_{e}$, $(0.54\pm0.38)\times 10^{-5}$~\cite{BESIII:2025biy}.

\begin{figure}[htbp]
  \centering
  \includegraphics[width=0.5\textwidth]{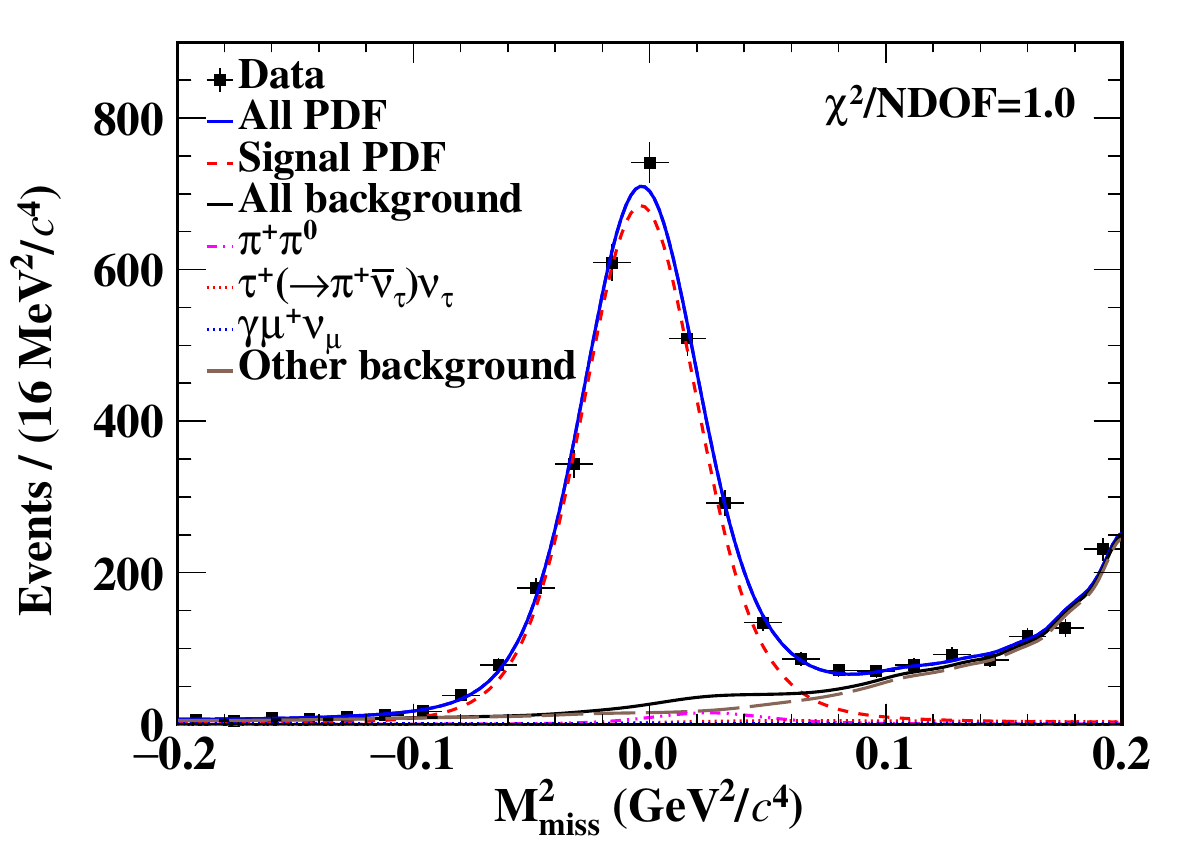}
  \caption{Fit to the $M_{\rm miss}^{2}$ distribution of the accepted candidates for $D^+\to \mu^+\nu_\mu$.
}
  \label{fig:crfit}
\end{figure}

To obtain the signal yield of $D^+\to\mu^+\nu_\mu$, we perform a fit to the $M_{\rm miss}^2$ distribution of the $D^+\to\mu^+\nu_\mu$ candidates in data. In the fit, the signal shape is modeled by the MC simulated shape convolved with a normal distribution with free parameters. The shapes of the peaking backgrounds from $D^+\to \tau^+(\to \pi^+\bar\nu_\tau)\nu_\tau$, $D^+\to\pi^+\pi^0$, $D^+\to\gamma\mu^+\nu_{\mu}$ and the remaining background are modeled by individual MC simulated events. The corresponding (probability density functions) PDFs are derived from individual simulated shapes with kernel estimation method~\cite{CRANMER2001198}. The yields of  $D^+\to \tau^+(\to \pi^+\bar\nu_\tau)\nu_\tau$ and $D^+\to\pi^+\pi^0$, corrected by the differences in misidentifying $\pi^+$ as $\mu^+$ between data and MC simulation, as well as $D^+\to\gamma\mu^+\nu_{\mu}$ are fixed in the fit, while the yields of signal and the remaining background are floated.
The fit result is shown in Fig.~\ref{fig:crfit}. The mean and resolution of the smeared normal function are -2 MeV$^2$/$c^4$ and 11 MeV$^2$/$c^4$, respectively. From this fit, we obtain the signal yield of $D^+\to \mu^+\nu_\mu$ to be $N_{\rm DT}=2832.7\pm56.8$.
Consequently, the BF of $D^+\to\mu^+\nu_\mu$ is found to be
${\mathcal{B}}_{D^+\to\mu^+\nu_\mu}=(4.034\pm 0.080_{\rm stat})\times10^{-4}$.

\begin{table}[htbp]\centering
  \caption{Systematic uncertainties in the BF measurement.}
  \label{tab:sys_crs}
  \begin{tabular}{lc}
\hline
\hline
  Source & Uncertaintiy (\%) \\
\hline
  $M_{\rm BC}$ fit                                           &0.30   \\
  $\mu^+$ tracking                                          & 0.06    \\
  $\mu^+$ PID                                                & 0.10    \\
  $E_{\rm max}^{{\rm extra~}\gamma}$ \&\& $N_{\rm charge}^{\rm extra}$  & 0.08   \\
  $M_{\rm miss}^{2}$ fit                                  & 0.90  \\
  Effect of $D^+\to \gamma \mu^+\nu_\mu$  &  0.17   \\
  MC statistics                                                & 0.16    \\
  \hline
  Total & 0.99\\
  \hline
  \hline
  \end{tabular}
\end{table}

Table \ref{tab:sys_crs} summarizes the sources of systematic uncertainties in the measurement of the BF of $D^+\to\mu^+\nu_\mu$. The systematic uncertainty in the $M_{\rm BC}$ fit is estimated with alternative signal and background shapes.
The alternative signal shapes are obtained by varying the parameters of the smeared normal distributions by $\pm 1\sigma$. The alternative background shape is obtained by varying the endpoint of the ARGUS function by $\pm$0.2 MeV/$c^2$. The relative difference of the ST yields between data and the inclusive MC sample $\mathcal{R}(N_{\rm data}/N_{\rm MC})$ is assigned as the systematic uncertainty.  Adding these systematic effects in quadrature gives a systematic uncertainty of 0.30\% due to the $M_{\rm BC}$ fit.

The $\mu^+$ tracking and PID efficiencies are studied with the control sample of $e^+e^-\to\gamma\mu^+\mu^-$ events.
After correcting the signal efficiency by $f^{\rm cor}_{\mu\,\rm PID}$, we assign 0.06\% and 0.10\%
as the uncertainties in the $\mu^+$ tracking and PID, respectively.

The efficiency for the $E^{\mathrm{extra}~\gamma}_{\rm max}$ requirement is studied with a control sample of DT hadronic events; i.e., events where both $D^+$ and $D^-$ decay to one of the eight ST hadronic final sates.
The systematic uncertainty is taken to
be 0.08\% considering the efficiency differences between data and MC simulation.

The systematic uncertainty associated with the signal shape in the $M_{\rm miss}^2$ fit is estimated by using an alternative signal shape represented by a double normal distribution. The relative change of the signal yield, 0.89\%, is assigned as the systematic uncertainty from this source.
The systematic uncertainty due to the peaking  background is estimated by varying  the world average BFs of the two background components within $\pm 1\sigma$~\cite{PDG2024}. The larger relative change of the fitted signal yield, 0.06\% and 0.12\%, is assigned as the systematic uncertainty associated with the estimated yields of $D^+\to \tau^+\nu_{\tau}$ and $D^+\to \pi^+\pi^0$, respectively.
Additionally, the kernel bandwidth in the kernel density estimation of the combinatorial background PDF is varied through $0.5$, $1.0$, $1.5$, and $2.0$. The largest difference in the fitted signal yield, 0.10\%, is taken as a systematic uncertainty.
The total systematic uncertainty arising from the $M_{\rm miss}^2$ fit is determined to be 0.90\% by adding the four individual uncertainties in quadrature.

To consider the systematic uncertainty of the background of $D^+\to\gamma\mu^+\nu_\mu$, we vary the fixed yield of $D^+\to\gamma\mu^+\nu_\mu$ in the $M_{\rm miss}^2$ fit within $\pm 1\sigma$. The larger change of the measured BF, 0.17\%, is taken as the systematic uncertainty.
The systematic uncertainty arising from the limited MC sample size, including both ST and DT MC samples,
is 0.16\%.

Assuming all components are independent, the relative total systematic uncertainty in the BF measurement is determined to be 0.99\% by adding the contributions in quadrature. Accounting for this, the BF of $D^+\to\mu^+\nu_\mu$ is measured to be $(4.034\pm0.080_{\rm stat}\pm 0.040_{\rm syst})\times10^{-4}$.

To determine $f_{D^+}|V_{cd}|$, radiative corrections for ${\mathcal B}(D^+\to \mu^+\nu_\mu)$ are necessary as stated by Ref.~\cite{Bazavov:2017lyh}. The $D^+\to\gamma\ell^+\nu_\ell$ decay rate, partially due to structure-dependent bremsstrahlung was predicted to be in a large range of $(0.46\sim8.2)\times10^{-5}$~\cite{Burdman:1994ip,doi:10.1142/S021773230000267X,LU200375,doi:10.1142/S0217751X09043262,PhysRevD.61.114510,YANG2014778,YANG2017301}, but was recently determined to be $(0.54\pm0.38)\times 10^{-5}$ by BESIII~\cite{BESIII:2025biy}. The BESIII result gives its fraction of $(0.23\pm0.17)$\%, which has been fixed in the signal extraction. The known short-distance electroweak correction increases ${\mathcal B}(D^+\to \mu^+\nu_\mu)$ by 1.8\%~\cite{Bazavov:2017lyh,SIRLIN198283} and the long-distance electroweak correction lowers ${\mathcal B}(D^+\to \mu^+\nu_\mu)$ by 2.5\%~\cite{Kinoshita:1959ha}, with an uncertainty of 0.6\% due to unknown electromagnetic corrections that depend upon the mesons’ structure~\cite{Bazavov:2017lyh}.
Combining our BF with
the world average values of $G_F$, $m_{\mu^+}$, $m_{D^+}$ and
the $D^+$ lifetime $\tau_{D^+}=(1.033\pm0.005)\times10^{-12}$~\cite{PDG2024} in $\Gamma_{D^+\to\mu^+\nu_\mu}^{(0)}$ and incorporating above corrections, we obtain
\begin{equation}
f_{D^+}|V_{cd}|=(48.02\pm0.48_{\rm stat}\pm0.24_{\rm syst}\pm0.12_{\rm input}\pm0.15_{\rm EM})~\mathrm{MeV}.
\nonumber
\end{equation}
Here the third(forth) uncertainty mainly arises from 0.2\%(0.3\%) uncertainty in $\tau_{D^+}$(radiation corrections).
Taking the magnitude of the $c\to d$ CKM matrix element $|V_{cd}|=0.22487\pm0.00068$ from the global SM fit
~\cite{PDG2024} we obtain
\begin{equation}f_{D^+}=(213.5\pm2.1_{\rm stat}\pm1.1_{\rm syst}\pm0.8_{\rm input}\pm0.7_{\rm EM})~\mathrm{MeV}.
\nonumber
\end{equation}
 Alternatively, taking the averaged decay constant
$f_{D^+}=(212.0\pm0.7)~\mathrm{MeV}$ from recent LQCD calculations~\cite{FlavourLatticeAveragingGroupFLAG:2021npn} as input, we determine
\begin{equation}
|V_{cd}|=0.2265\pm0.0023_{\rm stat}\pm0.0011_{\rm syst}\pm0.0009_{\rm input}\pm0.0007_{\rm EM}.
\nonumber
\end{equation}
Here, the uncertainties due to the input values of $\tau_{D^+}$ and $|V_{cd}|$~($f_{D^+}$) are 0.2\% and 0.3\%~(0.3\%), respectively.

Using our measurement,
the ratio of $\mathcal{B}_{D^+\to\mu^+\nu_\mu}$
over the world average value of
$\mathcal{B}_{D^+\to\tau^+\nu_\tau}=(1.20\pm0.27)\times10^{-3}$~\cite{PDG2024} is determined to be
$\mathcal{R}_{\tau/\mu}=2.97\pm0.67,$
which agrees with the SM prediction of $2.66\pm0.01$ from Eq.~\ref{eq1} within uncertainties.

Finally, we measure the separate BFs of $D^+\to\mu^+\nu_\mu$ and $D^-\to\mu^-\bar{\nu}_\mu$ to be $(3.99\pm0.11_{\rm stat.} \pm 0.04_{\rm syst.})\times10^{-4}$ and $(4.07\pm0.11_{\rm stat.}\pm 0.04_{\rm syst.})\times10^{-4}$, respectively.
From these we determine the BF asymmetry to be
$A_{\rm CP}=\frac{{\mathcal B}_{D^+\to\mu^+\nu_\mu}-{\mathcal B}_{D^-\to\mu^-\bar{\nu}_\mu}}{{\mathcal B}_{D^+\to\mu^+\nu_\mu}+{\mathcal B}_{D^-\to\mu^-\bar{\nu}_\mu}}=(-1.0\pm2.0_{\rm stat.}\pm0.9_{\rm syst.})\%,$
where  systematic uncertainties are assigned to account for the uncorrelated contributions between the charge-conjugated modes, arising  from the $\mu^\pm$ tracking and PID,
the ST yields, the limited MC sample sizes, and the $M^2_{\rm miss}$ fits. The $A_{\rm CP}$ is
compatible with the null hypothesis.

In summary, using the $e^+e^-$ collision data sample corresponding to an
integrated luminosity of 20.3~fb$^{-1}$ collected at $E_{\rm c.m.}=3.773$~GeV with the BESIII detector,
we report the most precise measurements of the BF of $D^+\to\mu^+\nu_\mu$, the decay constant $f_{D^+}$, and the magnitude of the $c\to d$ CKM matrix element $|V_{cd}|$. The precision is improved by a factor of 2.3 compared to
the previous best measurement~\cite{BESIII:2013iro}. In addition, we have searched for LFU and $\!CP$ violation in  $D^+\to \ell^+\nu_\ell$ decays, yet no statistical significant violation has been observed.

The BESIII Collaboration thanks the staff of BEPCII and the IHEP computing center for their strong support. Authors appreciate for helpful discussions with Prof. Maozhi Yang and Prof. Yuming Wang. This work is supported in part by National Key R\&D Program of China under Contracts Nos. 2023YFA1606000, National Natural Science Foundation of China (NSFC) under Contracts Nos. 11635010, 11735014, 11935015, 11935016, 11935018, 12025502, 12035009, 12035013, 12061131003, 12192260, 12192261, 12192262, 12192263, 12192264, 12192265, 12221005, 12225509, 12235017, 12361141819; the Chinese Academy of Sciences (CAS) Large-Scale Scientific Facility Program; the CAS Center for Excellence in Particle Physics (CCEPP); Joint Large-Scale Scientific Facility Funds of the NSFC and CAS under Contract No. U2032104, U1832207;  the Excellent Youth Foundation of Henan Scientific Committee under Contract No. 242300421044; 100 Talents Program of CAS; The Institute of Nuclear and Particle Physics (INPAC) and Shanghai Key Laboratory for Particle Physics and Cosmology; German Research Foundation DFG under Contracts Nos. 455635585, FOR5327, GRK 2149; Istituto Nazionale di Fisica Nucleare, Italy; Ministry of Development of Turkey under Contract No. DPT2006K-120470; National Research Foundation of Korea under Contract No. NRF-2022R1A2C1092335; National Science and Technology fund of Mongolia; National Science Research and Innovation Fund (NSRF) via the Program Management Unit for Human Resources \& Institutional Development, Research and Innovation of Thailand under Contract No. B16F640076; Polish National Science Centre under Contract No. 2019/35/O/ST2/02907; The Swedish Research Council; U.S. Department of Energy under Contract No. DE-FG02-05ER41374.

\bibliography{bibliography}

\end{document}